\documentclass[9pt,twocolumn,twoside]{osajnl}
\usepackage[utf8]{inputenc}
\DeclareUnicodeCharacter{2009}{\,}

\journal{optica} 

\setboolean{shortarticle}{true}

\title{Regenerative Terahertz Quantum Detectors}

\author[1,$\dagger$]{Paolo Micheletti}
\author[1,*]{Jerome Faist}
\author[1]{Tudor Olariu}
\author[1]{Urban Senica}
\author[1]{Mattias Beck}
\author[1,]{Giacomo Scalari}

\affil[1]{Institute for Quantum Electronics, ETH Zurich, 8093 Z\"urich, Switzerland}
\affil[$\dagger$]{e-mail: pmicheletti@ethz.ch}

\affil[*]{Corresponding author: jfaist@ethz.ch}

\ociscodes{(140.0140) Lasers and laser optics; (140.3070) Infrared and far-infrared lasers; (140.5965) Semiconductor lasers, quantum cascade.}

\begin{abstract}
Because of the ultrafast and photon-driven nature of the transport in their active region, we demonstrate that quantum cascade lasers can be operated as resonantly amplified terahertz detectors. Tunable responsivities up to 50 V/W and noise equivalent powers down to 100 pW/Hz$^{1/2}$ are demonstrated at 4.7 THz. Constant peak responsivities with respect to the detector temperature are observed up to 80K. Thanks to the $\approx$ ps intersubband  lifetime electrical bandwidths larger than 20 GHz can be obtained, allowing the detection of optical beatnotes from quantum cascade THz frequency combs. 
\end{abstract}

\setboolean{displaycopyright}{true}

\begin{document}
	
\maketitle
	
\section{Introduction}

Recent years saw an impressive development in the field of high speed/ high frequency devices operating at THz wavelengths. Specifically, the demonstration and development of quantum cascade laser frequency combs first in the Mid-ir~\cite{Hugi:2012ep} and then in the THz~\cite{Burghoff:2014hpa} is having major impact in the field of spectroscopy~\cite{villares_dual-comb_2015} and holds high promise also for other areas such as telecom for the next 6G and local area networks~\cite{Harter_Koos_NatPhot2018}. As the development of THz QCL comb sources~\cite{rosch_heterogeneous_2018,forrer_photon-driven_2020,Digasparenanophot2020} and their understanding~\cite{opacak_theory_2019,BurghoffOptica2020} proceeds at a fast pace, the call for ultrafast, highly responsive THz detectors is imperative. Recent advances using graphene~\cite{VitiNanoLett2020}  and Quantum Well Infared Photodetectors (QWIPs) inserted in nanoresonators~\cite{JeanninAPL2020} are prominsing but a highly responsive, ultrafast, high temperature, broadband THz detector is still lacking. 
We propose here an implementation of the concept of regenerative amplification for the detection of THz radiation. We leverage on the ultrafast dynamics of the quantum cascade active medium driven to perform resonant detection. In a regenerative amplifier, a very large overall signal gain is achieved by positive feedback onto the input of the amplifier. Initially developed and used in early radios~\cite{hong2004history}, this approach was also considered for optical systems since a laser below threshold can be seen as a regenerative amplifier~\cite{siegman_lasers_1986-1}. Low-light signal detection using regenerative amplification was nevertheless investigated experimentally~\cite{LITTLER:1990vz,LITTLER:1992iz}. The key disadvantage of regenerative amplification is its narrow bandwidth, which is the reason why it is employed in optical systems mostly for amplification of short pulses~\cite{SQUIER:1991te} and not for the detection of optical signals. Indeed, in the visible and near infrared photodiodes combining very large quantum efficiencies ($\eta > 0.5$) and low dark currents, combined with low noise electronics, enable extremely sensitive optical receivers to be realized. 

In contrast, in the THz this approach is much less attractive since the photon energy is smaller than $k_B T$ at room temperature. Indeed, quantum cascade detectors~\cite{Graf:2004gc} and quantum well infrared photoconductors~\cite{Liu:2004ef} operating in the THz have been reported, with high frequency capabilities\cite{Li:2017ht}. Nevertheless, and in contrast to the situation in the mid-infrared where  the use of optimal doping~\cite{Hao:2014ko} combined with microcavities~\cite{Palaferri:2018ia} enabled the operation up to room temperature, the THz QCDs~\cite{JeanninAPL2020} operate
below liquid nitrogen temperatures, and excellent performances are typically obtained at 4 K. The situation is even more extreme in superconductor hot electron bolometers where very low NEPs are achieved at millikelvin temperatures. \cite{Shurakov_2015}.

Quantum cascade lasers based on intersubband transition have enabled the generation of coherent radiation in the mid-infrared and the development of sensing applications because of their capability to operate at room temperature. The recent improvement of terahertz quantum cascade lasers, based on the simultaneous use of larger band discontinuities combined with a simple two-well quantum cascade structure and numerical optimization, has brought the operation temperature of these devices in the range of thermoelectric (Peltier) coolers~\cite{Franckie:2020gb,Bosco:2019ea,khalatpour_high-power_2020}. 

One unique property of quantum cascade lasers is their ultrafast photon-driven transport, arising naturally from the combination of tunneling electron injection and an ultrashort upper state lifetime. As a result, the voltage-current characteristic of the device shows a sharp conductance discontinuity at threshold as the photon-driven transport is added to the non-radiative current~\cite{Sirtori:1998p254}. The ultrafast nature of the response to that current is responsible for the very strong beatnote appearing at the round trip frequency observed in quantum cascade laser frequency combs; in this situation the laser is detecting its own light intensity modulation \cite{li_dynamics_2015}.

\begin{figure*}[!htb]
	\centering
	\includegraphics[width=\linewidth]{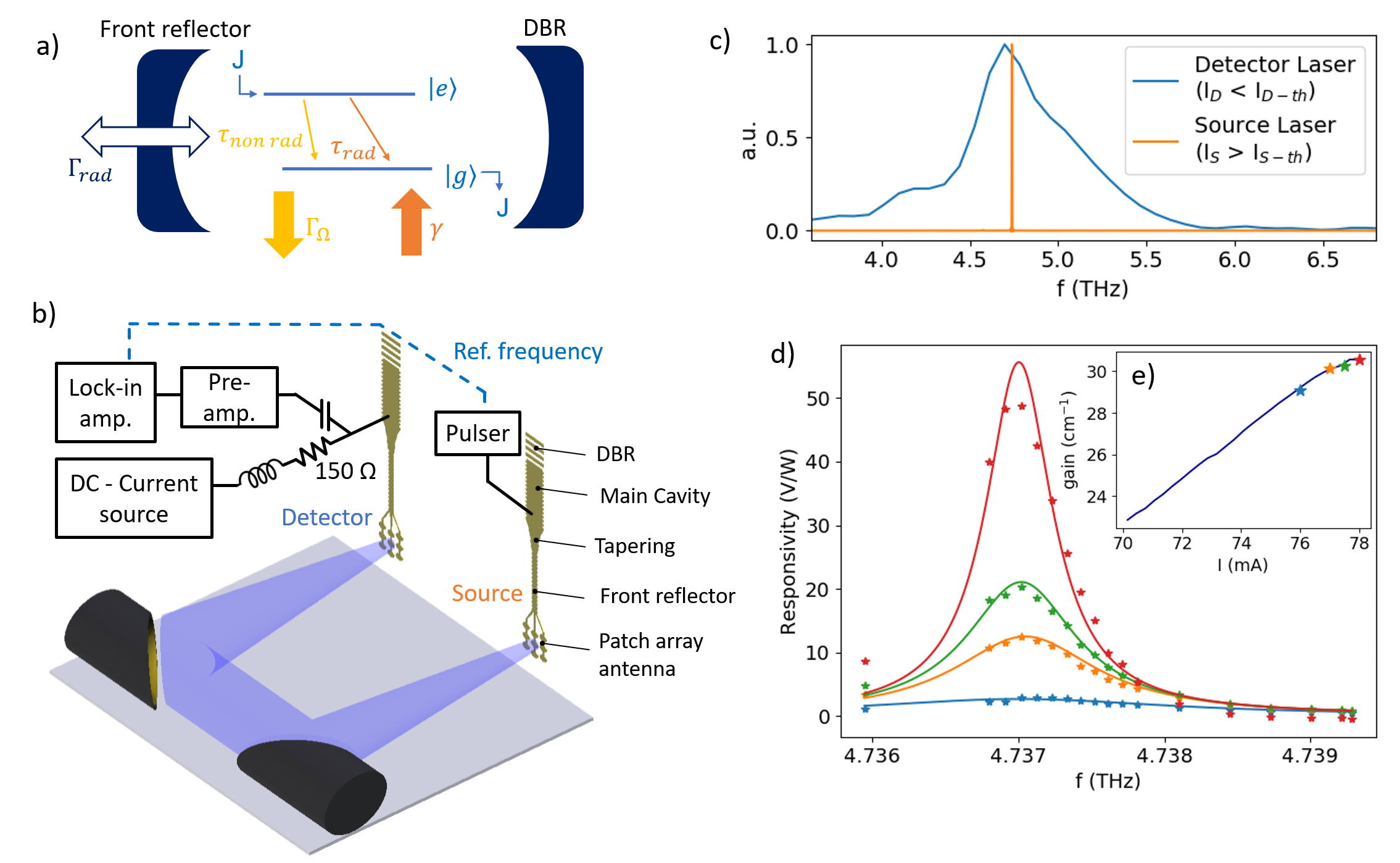}
	\caption{A resonantly amplified quantum cascade detector (RAMP-QCD). a) Schematic representation of the resonant cavity with the different terms entering equation for the responsivity \ref{equ:Resp_microcav}. b) Experimental setup: The source laser is driven in pulsed mode and its emission  focused onto the detector laser driven in continuous wave below threshold. The signal is detected on a  150 $\Omega$ load resistor placed in series with the laser. The different components of the patch-array antenna-single mode QCL are indicated on the representation of the source laser. c) Spectra of the luminescence of the detector laser below threshold (blue) and of the emission source laser above threshold (orange). d) Detector responsivity as a function of source frequency for different detector bias currents.  For each current the experimental data are fitted with equation \ref{equ:Iph}. The values of gain resulting from the fit are reported in e) as a function of the detector bias current. The points corresponding to the curves reported in d) are indicated by the colored $\star$. 
	}
	\label{fig:intro}
\end{figure*}

In the exact same way as a device can be detecting its own radiation above threshold, we anticipate that a photocurrent can be induced in a quantum cascade laser driven {\em below} threshold but such that the active region, containing $N_p$ periods has a gain of $g$ at the photon energy  $ h \nu$. The photocurrent generated in a section $\delta z$ of a Fabry-Perot laser cavity ($\delta I_{ph}$) will be equal to  
\begin{equation}
\delta I_{ph} =  \frac{P_{circ}(z)}{N_p} \frac{e}{h \nu} g \delta z
\label{equ:dI-rad}
\end{equation}
Where $P_{circ}(z) = |E_{circ}(z)|^2 $ is the circulating power inside the cavity.
The total photocurrent can be computed by taking explicitly the spatial dependence of the field inside the laser cavity, which can be expressed as a function of the net gain $g_{net} = g - \alpha_w $ (where $g$ is the gain and $\alpha_w$ the waveguide losses), and amplitude mirror reflectivities $r_1$, $r_2$ for every wavelength $\lambda$~\cite{siegman_lasers_1986-1}:

\begin{equation}
E_{circ}(z) = \frac{E_{in} \sqrt{\eta_{opt} (1-|r_1|^2)}}{1+r_1r_2e^{2(g_{net} -ik)L_{cav}}}\bigg[  e^{ - (g_{net}/2 + ik )z} + r_2e^{(g_{net}/2 + ik )z}	\bigg] 
\label{equ:Ecric}
\end{equation}

where $E_{in}$ is the incoming optical field and $k = 2 \pi/\lambda$. An optical efficiency $\eta_{opt}$ is introduced to take into account the fraction of incoming power effectively coupled into the laser cavity. 
Integrating $\delta I$ over the whole cavity length $L_{cav}$ yields the total photocurrent:

\begin{equation}
I_{ph} = \frac{P_{circ}(0)}{h \nu / q_0 N_p}\frac{g}{\alpha_w - g} \bigg[ e^{ - g_{net}L_{cav}} -1 +|r_2|^2 \left( 1- e^{  g_{net} L_{cav}} \right) \bigg] .
\label{equ:Iph}
\end{equation}

When considering a microcavity device, the spatial dependence of the field can the be neglected, allowing to write the responsivity $\mathcal{R}$ ( in A/W) on resonance as 
\begin{equation}
\mathcal{R} = \frac{1}{N_p} \frac{e}{h \nu} \frac{4 \gamma \Gamma_{rad}}{\left ( \Gamma_{rad} + \Gamma_{\Omega} - \gamma \right )^2 }
\label{equ:Resp_microcav}
\end{equation}

where the gain $\gamma$, the waveguide losses $\Gamma_\Omega$ and the radiative losses $\Gamma_{rad}$ are now expressed as rates~Fig.\ref{fig:intro} a). In contrast to conventional photodetectors, the dark current $i_d$ required  to maintain the gain is relatively large, such that the dynamical impedance of the device must be taken into account to evaluate the voltage responsivity. The dynamical impedance can be estimated in an analytical model by using the density matrix approach first developed by Kazarinov and Suris~\cite{Kazarinov:1972p1566} in the case of a superlattice and then extended to quantum cascade lasers~\cite{Sirtori:1998p254}. In this way, the current density $J$ flowing between the injector subband and the upper state detuned in energy by an amount $\hbar \Delta$ is proportional to their population difference $\Delta n$ :
\begin{equation}
J(\Delta, \Delta n) = \frac{ 2 e |\Omega_{21}|^2}{\tau_{\parallel}} \frac{\Delta n}{\Delta^2 + 1/\tau_{\parallel}^2}
\label{equ:KS-subbands}
\end{equation}
where $\tau_{\parallel}$ is the in-plane scattering lifetime and $\hbar \Omega_{21}$ the coupling between the subbands. When combined with a rate equation for the upper state and assuming a constant total electron density between the two subbands, the expression~\ref{equ:KS-subbands} yields for the current density for a given radiative current $J_{rad}$ from the upper state:
\begin{equation}
J(\Delta, J_{rad}) = \frac{2 e |\Omega_{21}|^2 \tau_{\parallel}}{1 + \Delta^2 \tau_{\parallel}^2 + 4 \Omega_{21}^2 \tau_{3} \tau_{\parallel}} \left ( n_s + \frac{2 \tau_3}{e} J_{rad} \right ),
\label{equ:KS-rad}
\end{equation}
where $\tau_3$ is the upper state population lifetime. Equation \ref{equ:KS-rad} reverts back to the well-known result at $J_{rad} = 0$. Using the above result, the differential photoresistance (per period) can be extracted by computing 
\begin{equation}
R_d = \frac{ \hbar}{e} \left . \frac{\partial \Delta}{\partial J_{rad}} \right |_{dJ = 0} = \frac{\hbar}{e^2 \Delta n_s} \frac{\tau_3}{\tau_{\parallel}^2} \left ( \Delta^2 \tau_{\parallel}^2 + 1+ 4 | \Omega_{21}|^2 \tau_3 \tau_{\parallel} \right )
\label{equ:Rd}
\end{equation}
This resistance is in general smaller than the differential resistance of the structure operated below threshold:
\begin{equation}
R_{\text{nr}} = \frac{ \hbar}{e} \left . \frac{\partial \Delta}{\partial J} \right |_{J_{rad} = 0} =  - \frac{\hbar}{e^2 \Delta n_s} \frac{\left ( \Delta^2 \tau_{\parallel}^2 + 1+ 4 | \Omega_{21}|^2 \tau_3 \tau_{\parallel} \right )^2}{4 |\Omega_{21}|^2 \tau_{\parallel}^3} 
\label{equ:Rnr}
\end{equation}
Indeed, the ratio of these two resistances is given by
\begin{equation}
\frac{R_d}{R_{\text{nr}}} = - \frac{ 4 | \Omega_{21}|^2 \tau_{\parallel} \tau_3}{\Delta^2 \tau_{\parallel}^2 + 1 + 4 | \Omega_{21} |^2 \tau_{\parallel} \tau_3}
\label{equ:Rd_Rnr}
\end{equation}
As the photovoltage is proportional to $R_d$ and the noise voltage (caused by shot noise) to $R_{\text{nr}}$, it is of course desirable to bring the above ratio as close to unity as possible. This is the case for thin injection barriers in the strong coupling limit defined by 
\begin{equation} 
4 | \Omega_{21} |^2 \tau_{\parallel} \tau_3 >>1
\end{equation}
This result enables us to write the noise equivalent power  of the device, assuming the noise is dominated by the shot noise:
\begin{equation}
\text{NEP} =\frac{ \sqrt{2 e i_{d} B} }{\mathcal{R}} \frac{R_{\text{nr}}}{R_d}
\label{equ:nep}
\end{equation}
where $i_d$ is the current and the noise is detected in a bandwidth $B$.

\section{Experimental Results}

An implementation of the Resonantly AMPlified Quantum Cascade Detector (RAMP-QCD) is achieved exploiting a patch-array antenna-single mode QCL~\cite{Bosco:2016dk}~Fig.\ref{fig:intro} b). The device consists of a double metal waveguide with a distributed Bragg reflector (DBR), acting as a high reflectivity mirror, and a first order distributed feedback (DFB) grating acting as a narrow-band front mirror. The two mirrors are designed with the help of finite element simulations performed with COMSOL which yielded a back mirror reflectivity of $\mathsf{R}_{back}$ = 0.95~\cite{bosco2020thz} and a front mirror reflectivity of $\mathsf{R}_{front}$ = 0.85 with a bandwidth of $\approx 100 GHz$ [ supporting material section 1 (SM1) ]. A patch-array antenna, integrated with the top metal contact, provides for efficient light out-coupling. At the same time it eases the in-coupling of light into the double metal waveguide, thus making the device suitable to work both as laser and RAMP-QCD. An in-coupling efficiency of 18 $\%$ is estimated with a a full-wave 3D numerical simulation performed with the software CST~[SM1].

\begin{figure}[!htb]
	\centering
	\includegraphics[width=\linewidth]{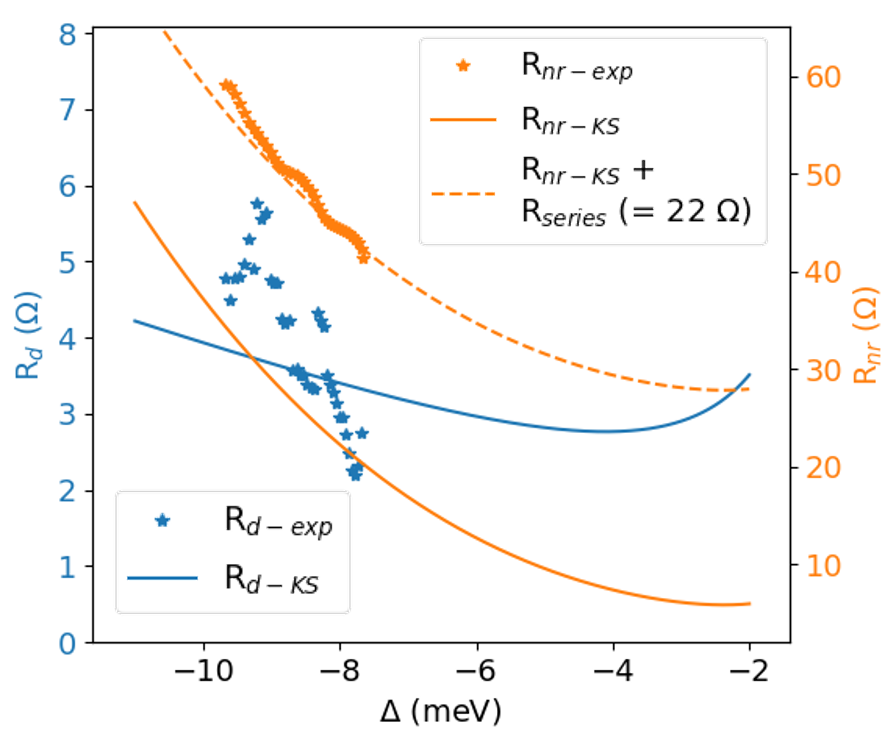}
	\caption{Non-radiative differential resistance ($R_{NR}$) and photoresistance ($R_d$) as a function of energy detuning $\Delta$. The orange $\star$ represent the experimental non-radiative resistances ($R_{nr - exp}$) measured in CW at a temperature of 20 K. The non-radiative resistance calculation based on the Kazarinov-Suris model (Eq. \ref{equ:Rnr}) is depicted with the orange solid line, the dashed line represents the same quantity with the addition of the series resistance (R$_{series} = 22$ $  \Omega$). The photoresistance is reported in blue. The $\star$ indicate the values resulting from the fitting of the experimental measurement of responsivity with equation \ref{equ:Iph} ($R_{d - exp}$), while the solid line indicates the calculation based on the Kazarinov-Suris model $R_{d - KS}$ (Eq. \ref{equ:Rd}).  }
	\label{fig:Rd_Rnr}
\end{figure}

\begin{figure*}[!htb]
	\centering
	\includegraphics[width=\linewidth]{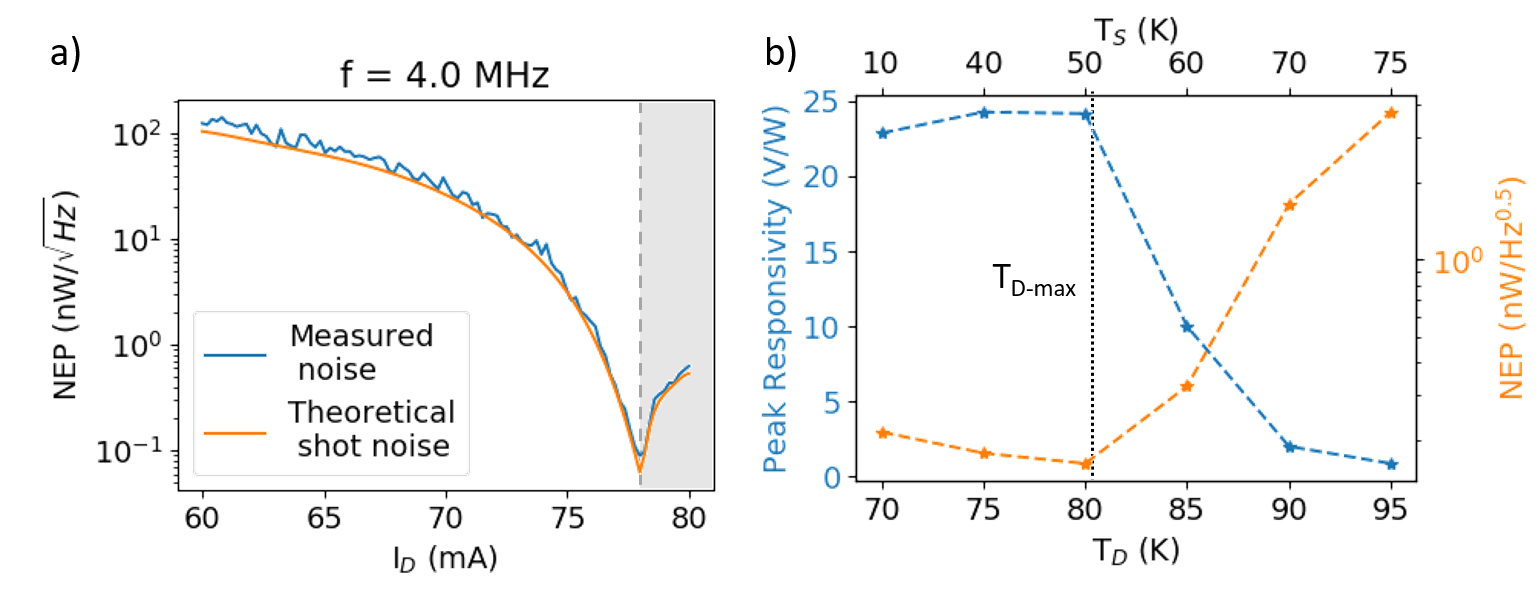}
	\caption{NEP in function of detector bias current a). The NEP, computed as $NEP = N/\mathcal{R}_{V/W}/R_{NR}$ is reported in blue. The power noise N is measured with a Rohde\&Schwarz FSW-67 spectrum analyzer connected to the detector load resistor. Noise Spectra from DC to 5 MHz are measured for different detector biases with a resolution bandwidth of 10 Hz and averaging over 10 samples. The results reported in the plot are relative to a frequency of 4 MHz and a device temperature of 15 K. The NEP obtained considering only the detector bias current shot noise ($N =  2 e I_{d} B R_{NR} $) is depicted in orange. The gray area represents the region above threshold. Peak responsivity in function of detector laser temperature b). The source laser is driven in pulsed mode with a fixed peak current (209 mA). For each temperature of the detector the source laser temperature is tuned to maximise the optical signal. The maximum responsivity reached by sweeping the detector bias is reported in blue. The NEP corresponding each responsivity and calculated from the bias current shot noise is reported in orange. The dashed line indicates the temperature at which the detector laser cannot reach threshold.}
	\label{fig:NEP_Td}
\end{figure*}

Two devices are used in the experiment, the emission of one laser, used as source, is focused onto the second, which is kept \textit{below threshold} and is used as detector (see experimental layout in~Fig.\ref{fig:intro} b). The spectra of the two lasers are centered around the same frequency (4.737 THz) as shown in Fig.\ref{fig:intro} c), where the luminescence of the detector laser is compared with the spectrum of the source laser above threshold. The optical resonant response of the detector is measured by temperature tuning the source laser in the range 10-55 K for fixed values of the detector laser current. As the detector bias is increased but maintained below threshold, the gain of the structure grows, resulting in a higher responsivity and a narrower optical bandwidth. For current values above threshold the radiation emitted by the active region of the detector rapidly overcomes the incoming one. As shown in Fig.~\ref{fig:intro} d) where the computed responsivity is reported as a function of optical frequency, the resonant nature of the detector is clearly apparent. This resonant behaviour can be described with good accuracy by our model where $\mathcal{R}_{V/W} = I_{ph} R_d/P_{in}$ and $I_{ph}$ is computed from Eqn.~\ref{equ:Iph}, assuming values of parameters described in detail in~[SM1]. In particular, the values of optical gain assumed to compute the responsivities as a function of bias current are shown in Fig.~\ref{fig:intro} e). The optical efficiency $\eta_{opt}= 12 \%$ is deduced from a comparison of the slope efficiency of the laser with the theoretical value. For each value of the injected current, the fit also yields a value of the differential photoresistance in the range $R_d = 2-5$ $\Omega$.  This value is reported in Fig.~\ref{fig:Rd_Rnr} and can be compared to the expected one derived from the Kazarinov and Suris model of transport \ref{equ:Rd}. Firstly, a Schroedinger-Poisson self-consistent solver is used to convert the applied bias into a detuning $\Delta$. The computed coupling between the injector and the upper state wavefunction $\hbar\Omega_{21} = 0.8$ meV was adjusted down to $\hbar\Omega_{21} = 0.45$ meV to obtain a better fit to the data. We also obtained the electronic lifetimes $\tau_{||} = 0.31 ps$ and $\tau_2 = 4.7ps$ from a measurement of the device electroluminescence linewidth (~Fig. \ref{fig:intro} c) and maximum current density respectively. As shown in Fig.~\ref{fig:Rd_Rnr}, while the experimental differential resistance of the device $R_{nr}$ agrees well with the computed one if one assumes an additional series resistance of 22 $\Omega$~[SM2]. The observed differential resistance $R_d$ shows a much stronger field dependence than predicted in our simple model, probably reflecting the well-known limitations of this simplified approach. This analysis, however, points towards the relatively low ratio $ R_d/R_{NR} \approx $ 8 $\%$ as strongly limiting the device's performance. Increasing this value could be achieved by using a stronger injection $4|\Omega_{21}|^2\tau_{||}\tau_3 $ above the value of 3 used in this device.

\begin{figure*}[!htb]
	\centering
	\includegraphics[width=\linewidth]{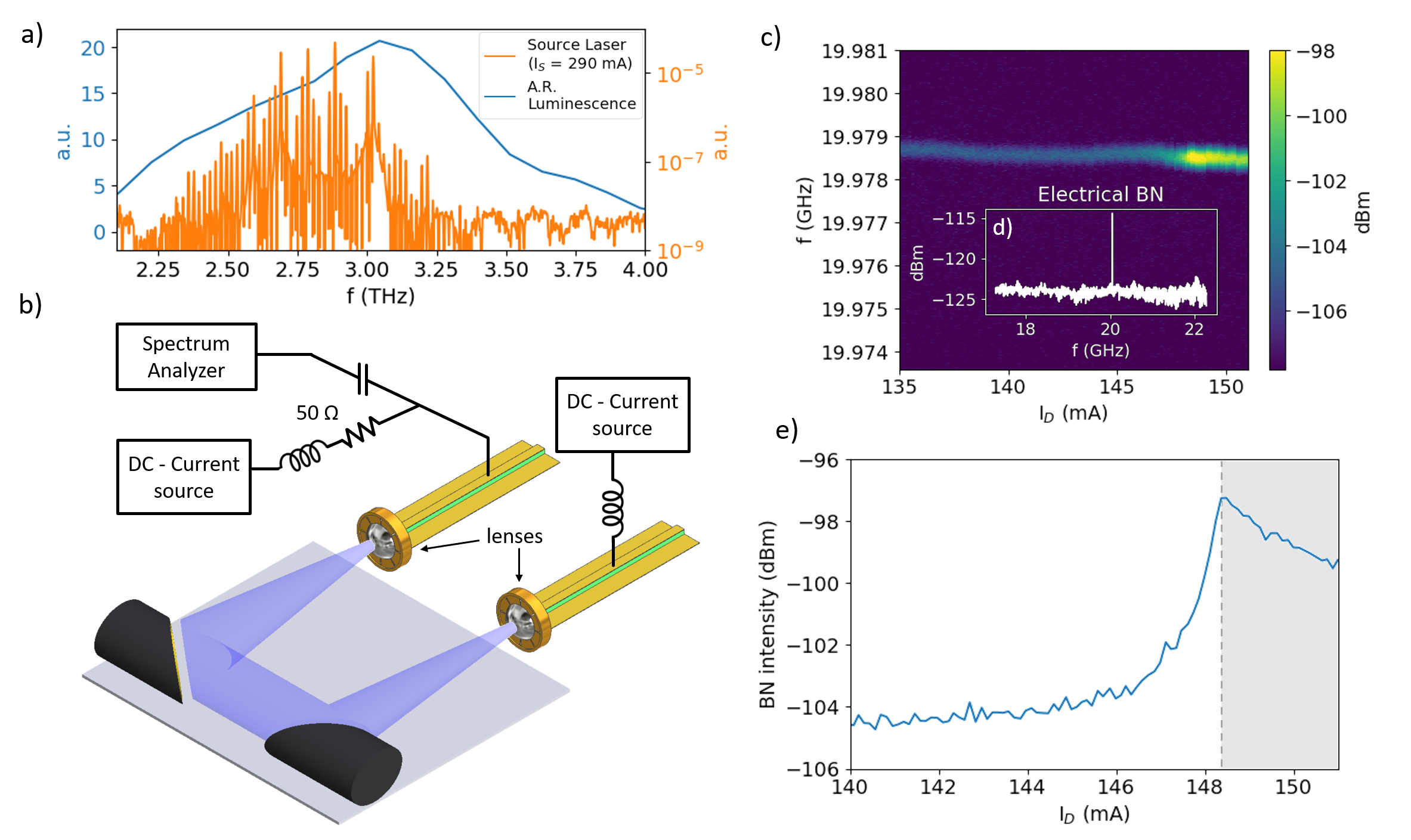}
	\caption{Optical detection of beatnote at GHz frequencies. (a) Comparison between the Active Region (AR) electroluminescence  and spectrum of the source laser. The AR luminescence (blue trace) is measured on a 0.345 mm long and 70 $\mu$m wide non-lasing device biased at 8V and kept at 10 K~\cite{forrer_photon-driven_2020}. The source laser spectrum  (orange trace) is measured with a Bruker Vertex 80 FTIR. The laser is kept at 15 K and driven CW at I$_S$ = 290 mA. Schematic of the experimental set-up b). The source and detector laser are placed on the foci of the parabolic mirrors. Both lasers are driven in CW, a 50 $\Omega$ load resistor is placed in series with the current source of the detector laser. A Rohde\&Schwarz FSW-67 spectrum analyzer is also connected to the detector QCL through a bias-tee. Optical beatnote map as a function of detector bias current c). The source laser bias is fixed at 290 mA and kept at 15 K. The detector laser temperature is 2 5K and its bias is swept from 135 mA to 151 mA. The spectrum analyzer is operated with a resolution bandwidth of 10 kHz. For each current the measurement is averaged on 20 samples. The electrical beat note measured on the bias tee of the source laser is reported in the inset. Maximum value of the beatnote intensity as a function of detector current e). The grey area indicates the region above threshold which is, in this configuration, I$_{D-th}$ = 148.4 mA.}
	\label{fig:lensed}
\end{figure*}

The noise of the detector is evaluated connecting a spectrum analyzer to the detector under bias while keeping the source laser off. From the measurement of the noise the minimum NEP is calculated using the responsivity obtained in the resonant case (~Fig. \ref{fig:NEP_Td} a), i.e. when the source and detector laser frequencies match. Values smaller than 100 pW/Hz$^{1/2}$ can be reached close to threshold. A good agreement is obtained with the theoretical NEP computed considering only the bias current shot noise, showing the shot noise to be effectively the dominant source of noise of the detector. As the signal is proportional to $R_d$ and the noise to $R_{NR}$, an improvement of this ratio would also benefit the NEP. A key feature of this approach to detect THz radiation is that the peak responsivity does not depend on the detector's temperature as long as it is below the maximum operating temperature of the laser. The decrease of upper state lifetime due to temperature increase indeed only shifts the threshold to higher values, resulting in a slightly larger noise level. As a result, the detector performances are expected to show only a weak temperature dependence.  

The detector's temperature dependence is investigated experimentally and shown in Fig. \ref{fig:NEP_Td} b).  In this measurement, a detector laser with a slightly higher central frequency is used such that the frequency matching between source and detector QCL is obtained when the temperature of the source is smaller than the one of the detector. The result shows that the peak responsivity stays approximately constant until 80 K (T$_{D-max}$), temperature at which the detector cannot reach its threshold anymore~[SM3]. Below T$_{D-max}$ the peak responsivity lies in the 20 - 25 V/W range. This value is comparable with the one obtained with the detector laser used in the previous experiment. Above the maximum operating temperature, the maximum gain of the active region starts to decrease resulting in a reduced peak responsivity. The NEP relative to each point is computed from the bias current shot noise. Despite the fact that at higher temperatures the maximum responsivity is obtained at higher currents, due to the increase of the threshold current, the NEP does not increase dramatically, thanks to the decrease of non radiative resistance with temperature. Above T$_{D-max}$ the decrease of gain, hence of responsivity, causes a substantial increase in NEP. 

Another fundamental advantage of exploiting QCLs for THz radiation detection is that, thanks to the ultrafast nature of the photon-driven transport in a QCL, the response of the RAMP-QCD can be extremely fast. The fact that electrical beatnotes of the order of several tens of GHz can be observed in THz QCL frequency combs~\cite{burghoff_terahertz_2014,rosch_-chip_2016, forrer_photon-driven_2020,ForrerAPL2021Harmonic}, suggests indeed that the cut off frequency of the RAMP-QCD response should lie in the same range. The easiest way to verify this hypothesis is to optically measure the beatnote of a THz frequency comb employing the RAMP-QCD. The device presented so far is strictly single-mode above threshold, therefore it is not suitable for this purpose. For this reason a pair of lasers based on a broadband active region~\cite{forrer_photon-driven_2020} is used, whose emission is centered around 3 THz (Fig. \ref{fig:lensed} a) blue trace). The devices consist of a double metal waveguide ridge 2 mm long and 40 $\mu$m wide. A silicon lens is placed on the front facet in order to lower its reflectivity and optimise the coupling inside/outside of  the ridge. The source laser is placed in one of the foci of the setup and biased in CW at $I_S$ = 290 mA. At this current the laser is in a comb state, displaying a wide spectrum with a bandwidth of $\approx$ 900 GHz and a single electrical beatnote at $\approx$ 20 GHz (~Fig \ref{fig:lensed} a) orange trace, d). The electroluminescence of the detector (blue trace, Fig \ref{fig:lensed} a) overlaps very well with the laser's signal. The detector is placed in front of the other mirror and connected to a spectrum analyzer through a bias-tee, and always biased below lasing threshold. In this configuration a clear optical beatnote can be observed on the spectrum analyzer for different bias of the detector laser (Fig \ref{fig:lensed} c), thus confirming that the electrical bandwidth of the detector is as high as 20 GHz. We would like to underline that we are not detecting the heterodyne signal of the incoming laser signal with the detector coherent field, as the detector is always below threshold. The detected signal is the  optical beating of the modes of the comb inside the detector's cavity, amplified by the detector's gain. The increase of the gain as the detector current is brought towards the threshold is also reflected into an increase of the optical beatnote intensity (Fig \ref{fig:lensed} c,e). 

\section{Discussion and conclusion}

We presented the application of the concept of regenerative amplification to THz detection exploiting QCL active regions as gain medium. The ultrafast photon driven transport that characterises QCLs offers the possibility of reaching high responsivities, wide electrical bandwidths and high operating temperatures.
A very promising outlook on future device architectures comes from the inspection of Eqn.\ref{equ:Resp_microcav}: a microcavity device would lead to responsivities up to 80 kV/W and NEP down to 1.5 pW/Hz$^{1/2}$ for a (10  x 10) $\mu$m$^2$ microresonator with a quality factor Q=20 and 2.5 kA/cm$^{-2}$ current density operating at 3.9 THz \cite{Bosco:2019ea}. Such active region characteristics correspond to the high performance 2-well operating up to 210 K: in this perspective we would obtain a highly responsive device up to high temperatures. The inherent ultrafast bandwidth of the detection mechanism would then be fully exploited in a microcavity geometry that reduces to the minimum the capacitive and inductive parasitics. Studies using lenses-coupled broadband lasers proved the possibility of recording optical beatnotes, thus confirming that bandwidths as high as 20 GHz can be reached with our approach and very likely can be extended to much higher frequencies as electrical beatnotes from THz QCLs have been detected up to 50 GHz and higher~\cite{ForrerAPL2021Harmonic}. Regenerative amplification proves to be an effective method for detecting THz radiation and has the potential to overcome the existing technology.

\section*{Funding Information}
The authors gratefully acknowledge the financial support of the Swiss National Science Foundation (SNF) and from H2020 European Research Council Consolidator Grant (724344) (CHIC) . 

\section*{Acknowledgments}

\section*{Supplemental Documents}
\emph{Optica} authors may include supplemental documents with the primary manuscript. For details, see \href{http://www.opticsinfobase.org/submit/style/supplementary-materials-optica.cfm}{Supplementary Materials in Optica}. To reference the supplementary document, the statement ``See Supplement 1 for supporting content.'' should appear at the bottom of the manuscript (above the references).

\bibliography{bibtex-library,bibGS}

\end{document}